\documentclass[authoryear,review,12pt]{elsarticle}

\usepackage{amsfonts}
\usepackage{mathtools}
\usepackage{mathrsfs}
\usepackage{multirow}

\journal{International Journal of Solids and Structures}

\begin{document}

\begin{frontmatter}

\title{Thermal buckling of thin injection-molded FRP plates with fiber orientation varying over the thickness}

\author[DSM_address]{A. Gualdi}
\author[TUE_address]{A.A.F. van de Ven}
\author[DSM_address,TUE_address]{J.J.M. Slot\corref{correspondingauthor}}
\ead{j.j.m.slot@tue.nl}

\cortext[correspondingauthor]{Corresponding author}
\address[DSM_address]{DSM Materials Science Center, P.O. Box 1066, 6160BB, Geleen, The Netherlands}
\address[TUE_address]{Department of Mathematics and Computer Science, Eindhoven University of Technology, The Netherlands}

\begin{abstract}
The different thermo-elastic properties of glass fibers and polymer matrices can generate residual thermal stresses in injection-molded fiber-reinforced plastic (FRP) objects. During cooling from mold to room temperature, these stresses can be relaxed by large deformations resulting from an instability of the unwarped configuration (i.e., buckling). This article investigates the thermal buckling of thin FRP disks via an analytical formulation based on the F\"oppl-von K\'arm\'an theory. Expanding on our previous work, cylindrical orthotropy with material parameters varying over the disk thickness is assumed in order to account for thickness dependency of the glass fiber orientation distribution. A disk parameter generalizing the thermal anisotropy ratio for homogeneous orthotropic disks is introduced and its relation with the occurrence and periodicity of buckling is discussed. This is done for a skin-core-skin model, for which the core-to-total thickness ratio is defined. For fiber orientation distributions typical of injection-molded disks, it is found that there exists a value of the thickness ratio for which no buckling occurs. It is also demonstrated that the periodicity of the first buckling mode is described by the generalized thermal anisotropy ratio, thus extending the results obtained for a homogeneous fiber orientation distribution. Improvements in the accuracy of the predictions for experimental data available in the literature when using the skin-core-skin model are shown. Finally, we study the relation between buckling temperature and disk thickness and propose an expression for the dependence of the normalized buckling temperature on the thermal anisotropy ratio. Results of FEM simulations are used to validate the proposed expression, proving its applicability and accuracy.
\end{abstract}


\end{frontmatter}

\allowdisplaybreaks[1]

\section{Introduction}
\label{sec_introduction}
Fiber-reinforced plastics (FRPs) are a class of materials consisting of a thermoplastic matrix reinforced by short glass fibers. They are often processed by injecting the molten polymer and the glass fibers into a mold (injection molding). The specific process parameters determine the orientation of the fibers in the final product, which in turn translates to its thermo-elastic properties (see for example \citet{Tseng1994}). As the injection-molded part cools down, it tends to shrink while building up stresses (\citet{Adhikari2016}). The problem of plates with residual thermal stresses has been widely investigated (see \citet{Hetnarski2014} for a comprehensive review). Residual stresses can be relaxed by sudden large deformation, a phenomenon referred to as \textit{buckling}. In general, it is possible to identify a parameter describing the magnitude of the internal stresses. Buckling occurs when this parameter exceeds a certain critical value, the \textit{buckling threshold}.\\
In a previous work (\cite{Gualdi2021}), we studied the thermal buckling of thin plates with uniform fiber orientation distribution through the wall thickness. A framework for the buckling analysis based on the von K\'arm\'an theory with constitutive linearity in the thermo-elastic material behavior (Hooke-Duhamel model) was formulated. The framework was specified for the particular case of a free disk with cylindrical orthotropy and uniform thermo-elastic properties through its thickness. A general technique to determine post-buckling deflections based on a perturbation approach and the Fourier-Galerkin method was proposed. It was found that the periodicity of the first buckling mode for orthotropic disks is fully determined by the anisotropy ratio of the elastic moduli (or equivalently by the ratio of the coefficients of thermal expansion), in agreement with experimental evidence available in the literature (\citet{Kikuchi1994,Kikuchi1996a,Kikuchi1996b}). By averaging the thermo-elastic properties of the material over the disk thickness, the model can be applied to systems where the fiber orientation distribution is a function of the vertical coordinate. However, a more general formulation accounting for thickness-dependent properties is required to properly investigate their effects on the buckling behavior.\\
The orientation of the glass fibers in injection-molded products with thin walls has been extensively investigated. Regardless of the mold geometry, the fibers tend to align in the flow direction or perpendicular to it (transverse direction) during the injection process depending on their distance from the mold walls. For flat parts, the orientation can be investigated by taking cross-sections perpendicular to their mid-planes. It is reported in the literature that the fibers are generally parallel to the walls with predominant alignment along the transverse direction close to the center and along the flow direction close to the surface. This is shown for reference geometries in \cite{Oumer2012} (rectangular cavity), \cite{Lionetto2021}  (tensile bar), \cite{Hamanaka2017} (rectangular plate), and \cite{Kikuchi1996a} (disk), and for a complex geometry in \cite{Baranowski2019} (bearing part for automotive applications). Variations in the predominant orientation are often observed close to the lateral edges of the mold. This effect is highlighted for fiber-reinforced PA66 disks in \cite{Kikuchi1996a}, where it is shown to become stronger for thicker disks and remain secondary for thinner geometries. In this work, we will assume that the in-plane variations in the fiber orientation distribution can be neglected, i.e., we will consider geometries with a large lateral-dimension-over-thickness ratio. When this assumption is satisfied, the fiber orientation distribution results in a 2D orthotropic material model with parameters varying through the thickness (vertical direction $z$). The material properties in the in-plane directions of orthotropy at a given $z$ are chosen such that the macroscopic material behavior resulting from the complex in-plane distribution is reproduced. An equivalent description is that of a continuous sequence of infinitely-thin orthotropic layers $z=$ constant. In the following, we will refer to this representation as \textit{orthotropic stack} or \textit{stack model}, in contrast to \textit{orthotropic single layer} or \textit{single-layer model} when the material parameters are constant through the thickness. The latter yields the formulation of our previous work as recalled in the paragraph above. Even though the material properties in the stack model are allowed to vary arbitrarily though the thickness, they are usually approximated with piecewise constant functions. The \textit{skin-core-skin} three-layer model is one of the most used representations. More accurate descriptions can generally be obtained by adding a shell layer between the core and the skin, resulting in the five-layer \textit{skin-shell-core-shell-skin} model. Techniques to translate continuous fiber orientation distributions to such models can be found for example in \cite{Lionetto2021} and \cite{Hamanaka2017}. These stack models have been introduced here as approximations of injection-molded geometries, but they can be also used to describe exactly composite plates made of orthotropic layers. In fact, the general formulation provided in our previous work can be applied to such problems too.\\
The buckling behavior of products with material properties varying through their thickness has been widely studied for composite materials. Often, it is investigated how the buckling parameter depends on the layer properties, in order to limit the amount of warpage by specific design of the stack build-up. It is generally found via analytical considerations based on linear elasticity that the ratios between the thicknesses of the various layers are crucial in the determination of the onset of buckling. This is shown for example in \cite{Lopatin2008}, where a symmetrical sandwich panel with orthotropic core subjected to in-plane compression is considered. When the stack is made of orthotropic layers, also the ratio between the elastic moduli in the directions of orthotropy becomes a relevant parameter. Such result can be found again in \cite{Lopatin2008} but also in \cite{Wankhade2020}, where the buckling of composite plates with orthotropic laminae is addressed via numerical methods. While layer thickness and level of orthotropy can be easily tuned in laminated composites, in injection-molded products they are the result of the flow path and process conditions and thus harder to control. Nevertheless, they are expected to play analogous roles in determining the buckling characteristics of the product. The relation between layer thickness, level of orthotropy, and observed warpage upon cooling is discussed for injection-molded disks in \cite{Kikuchi1996b}. Based on experiments and numerical simulations, the authors correlate the critical buckling temperature to $\left< R\alpha \right>$, the ratio between the coefficients of thermal expansion in the radial and tangential directions averaged over the disk thickness. This parameter is in fact a combination of the layer thickness ratios and levels of anisotropy, and the two contributions cannot be analyzed separately with the proposed approach. A modified linear correlation between $\left< R\alpha \right>$ and the normalized buckling temperature is also proposed.\\
The present work investigates the relation between thickness-dependent fiber orientation distributions and buckling behavior in thin injection-molded disks via equivalent stack models. The effects of layer thickness and orthotropy level are investigated independently and summarized by a disk parameter $\delta$, which is a generalized thermal anisotropy ratio and plays an analogous role to that of $\left< R\alpha \right>$ in \cite{Kikuchi1996b}. The formulation and solution technique presented in our previous work (\cite{Gualdi2021}) is used. When comparing the results to those of the single-layer formulation, an improvement in the accuracy of the predictions is observed when using a skin-core-skin formulation. Analytical considerations are used to improve the modified linear correlation between averaged thermal anisotropy and buckling temperature proposed in \cite{Kikuchi1996b}. The range of validity of the new expression is extended and the fit with the available numerical results is improved.\\
The paper is organized as follows. The F\"oppl-von K\'arm\'an equations for a stack of orthotropic circular layers are derived in Section \ref{sec_system}. The solution technique is also recalled in the same section. In Section \ref{sec_results}, the role of the generalized thermal anisotropy ratio $\delta$ is discussed for a skin-core-skin geometry. Results are compared to those of the single-layer model from our previous work. Finally, the correlation between normalized buckling temperature and thermal anisotropy ratio is studied and an improved fitting function is proposed.
\section{The F\"oppl-von K\'arm\'an equations for continuous orthotropic disks}
\label{sec_system}
Let us consider a thin disk of thickness $H$ and radius $R$. Let the center of the coordinate system coincide with the center of the midplane of the disk. The radial, tangential, and normal (or vertical) directions are indicated with $r$, $\varphi$, and $z$, respectively. We assume that the material behavior is described by the Hooke-Duhamel law (linear thermo-elasticity) and that the thermo-elastic properties are symmetric with respect to the midplane of the plate. Thermal stresses are taken proportional to $\Theta = \Theta \left( z \right) = \Theta \left( -z \right)$, the (symmetric) temperature difference due to cooling down of the disk, during which the temperature decreases from its higher initial value $T_i$ of the hot mold to its final lower value $T_f$ (the colder room temperature). Since large deflections $w$ can occur during cooling, the mechanics of deformation is best described by the F\"oppl-von K\'arm\'an theory. A general coordinate-free formulation of the F\"oppl-von K\'arm\'an equations for thin anisotropic and inhomogeneous plates was derived in \citet{Gualdi2021}. A solution technique based on perturbation schemes and Galerkin-Fourier approximations was also presented. The equations were then specified for the particular case of a cylindrically orthotropic disk with constant thermo-elastic properties through the thickness. Here, we want to describe disks with thickness-dependent fiber orientation distributions. We will still assume that each cross-section $z=\text{constant}$ is cylindrically orthotropic, but we will allow the material properties to be functions of $z$. Under these assumptions, the F\"oppl-von K\'arm\'an equations take the same structure as the particular homogeneous case in \citet{Gualdi2021}:
\begin{equation}
    \begin{dcases}
    \tilde{\Delta}_1^2 w = \left[ w, \chi \right]~; \\
    \tilde{\Delta}_2^2 \chi = -\frac{1}{2}\left[ w,w \right]~.
    \end{dcases}
    \label{eq_fvk_system}
\end{equation}
The expressions of the differential operators $\tilde{\Delta}_1^2$, $\tilde{\Delta}_2^2$, and $[\cdot,\cdot]$ are given in Appendix \ref{app_A} and differ from the ones in \cite{Gualdi2021} only in the material-dependent coefficients. The same holds for the four boundary conditions of vanishing normal and shear resultants, bending moment, and generalized shear force. Their explicit expressions are given in Appendix \ref{app_A}. The unknown deflection $w$ and Airy stress function $\chi$ (both functions of $\rho \coloneqq r/R$ and $\varphi$) are dimensionless. They can be obtained from their dimensional forms $w_{\text{dim}}$ and $\chi_{\text{dim}}$ via the following relations:
\begin{equation}
    w = \sqrt{\frac{\bar{\bar{E}}H}{\bar{\bar{D}}}} w_{\text{dim}}~, \quad \chi = \frac{H}{\bar{\bar{D}}} \chi_{\text{dim}}~,
\end{equation}
where
\begin{equation}
    \bar{\bar{E}} = \frac{1}{H} \int_{-H/2}^{H/2} \tilde{E}(z) \, \mathrm{d}z~, \quad \bar{\bar{D}} = \int_{-H/2}^{H/2} \frac{\tilde{E}(z) z^2}{1-\tilde{\nu}(z)^2} \, \mathrm{d}z~. \label{eq_def_barbars}
\end{equation}
The previous relations and the differential operators in (\ref{eq_fvk_system}) involve geometrical averages and ratios of the thermo-elastic properties in the two directions of orthotropy. Their definitions are analogous to those in \cite{Gualdi2021} and are here just recalled:
\begin{gather}
    \tilde{E}(z) \coloneqq \sqrt{E_r(z) E_\varphi(z)}, \quad \tilde{\nu}(z) \coloneqq \sqrt{\nu_{r\varphi}(z) \nu_{\varphi r}(z)}~, \quad \tilde{k}(z) \coloneqq \sqrt{k_r(z) k_\varphi(z)}~; \\
    \omega(z) \coloneqq \sqrt{\frac{E_\varphi(z)}{E_r(z)}} = \sqrt{\frac{\nu_{\varphi r}(z)}{\nu_{r \varphi}(z)}}~, \quad \gamma(z) \coloneqq \sqrt{\frac{k_r(z)}{k_\varphi(z)}}~.
\end{gather}
For each value of $z$, $\omega = \gamma = 1$ represents isotropic behavior, $\omega, \gamma < 1$ corresponds to a predominant radial fiber orientation, and $\omega, \gamma > 1$ to a tangential one. As the present model is formulated in terms of the composite thermo-elastic properties, it can also be applied when anisotropic properties are the result of the presence of different reinforcements (e.g., platelets) and/or processing conditions (e.g., flow orientation).\\
Since the structure of the system and its boundary conditions are the same as for the more specific case described in \cite{Gualdi2021}, the same solution technique discussed there can be here applied directly. Therefore, $w$ and $\chi$ are sought for by expanding them in powers of the small parameter $\varepsilon$ as follows:
\begin{align}
    w(\rho,\varphi;\varepsilon) &= w^{(0)}(\rho,\varphi) + \varepsilon w^{(1)}(\rho,\varphi) + \varepsilon^2 w^{(2)}(\rho,\varphi) + ... ~;
    \label{eq_w_exp} \\
    \chi(\rho,\varphi;\varepsilon) &= \chi^{(0)}(\rho,\varphi) + \varepsilon \chi^{(1)}(\rho,\varphi) + \varepsilon^2 \chi^{(2)}(\rho,\varphi) + ... ~.
    \label{eq_chi_exp}
\end{align}
where the parameter $\varepsilon$ is defined as
\begin{equation}
    \varepsilon \coloneqq \pm \sqrt{\frac{\bar{\Theta}-\Theta_c}{\Theta_c}} = \pm \sqrt{\frac{\mu - \mu_c}{\mu_c}}~.
    \label{eq_def_eps}
\end{equation} 
Here, $\Theta_c$ is the minimum temperature difference required for buckling (still unknown) and $\bar{\Theta}$ is the average of $\Theta(z)$ over the thickness. The dimensionless buckling parameter $\mu$ and its critical value $\mu_c$ are defined by a more general version of Equation (40) from \cite{Gualdi2021}:
\begin{equation}
    \mu\left( \bar{\Theta} \right) = \frac{H^3}{\bar{\bar{D}}} \left( \frac{R}{H} \right)^2 \frac{k_{2,0} \left( a_0 + b_0 \right) - k_{1,0} \left( b_0 + c_0 \right)}{a_0 - c_0} \bar{\Theta}~, \quad \mu_c = \mu\left(\bar{\Theta}_c \right)~.
    \label{eq_def_mu}
\end{equation}
The coefficients $a_0$, $b_0$, $c_0$, $k_{1,0}$, and $k_{2,0}$ are functions of the thermo-elastic parameters  of the material only and their expressions are given in Appendix \ref{app_A}. For a temperature decrease, $\bar{\Theta}>0$. However, the sign of its normalization constant depends on the considered fiber orientation distribution and thus both positive and negative values of $\mu_c$ are of interest. We also define the dimensionless parameters $\lambda$ and $\delta$ as
\begin{equation}
    \lambda^2 \coloneqq \frac{c_0}{a_0}~, \quad \delta^2 \coloneqq \frac{k_{1,0}(b_0+c_0)}{k_{2,0}(a_0+b_0)}~. \label{eq_def_lambda_delta}
\end{equation}
As it will be demonstrated later, they will determine the buckling mode and its occurrence just like $\omega$ and $\gamma$ do in the uniform case.\\ 
After solving the system (\ref{eq_fvk_system}) order by order, the approximate solution is fully determined up to the second order and reads
\begin{gather}
	w(\rho,\varphi;\varepsilon) \approx \varepsilon w^{(1)}_N(\rho)\cos(\bar{m}\varphi)~; \\
	\chi(\rho,\varphi;\varepsilon) \approx \chi^{(0)}(\rho) + \varepsilon^2 \chi^{(2)}_N(\rho,\varphi)~.
\end{gather}
The complete expressions for the solutions at each order are given in Appendix \ref{app_B}.
\section{Results}
\label{sec_results}
Many commercial FEM software packages allow to specify thickness-dependent fiber orientation distribution tensors through a layer approach: the geometry is described as a stack of discrete layers, each characterized by a 2D material model. In the case of thin injection-molded parts, a commonly used discretization is the symmetric skin-core-skin geometry. This representation has proven successful to capture the mechanical behavior of parts with continuous through-the-thickness fiber orientation distributions while keeping the amount of required input parameters limited. In the following, we will thus focus on such three-layer geometry. The current model however is more general and can be applied to geometries with any number of layers.\\
Let $h$ be the thickness of the core layer and $(H-h)/2$ that of each skin layer ($H$ is thus the total disk thickness). We define a thickness ratio $\alpha$ as:
\begin{equation}
    \alpha \coloneqq \frac{h}{H} \in \left[ 0, 1 \right]~.
\end{equation}
The values of the thermo-elastic properties in each layer are constant and will be denoted with the subscripts $S$ or $C$ when referring to the skin or core layer, respectively. First, we will draw some general conclusions regarding the occurrence of buckling; then we will compare the predictions to those of the equivalent single-layer model; finally, the dependence of the buckling temperature on thickness and thermal anisotropy ratio will be discussed.

\subsection{Occurrence and periodicity of buckling}
\begin{figure}
    \centering
    \includegraphics[width=\linewidth]{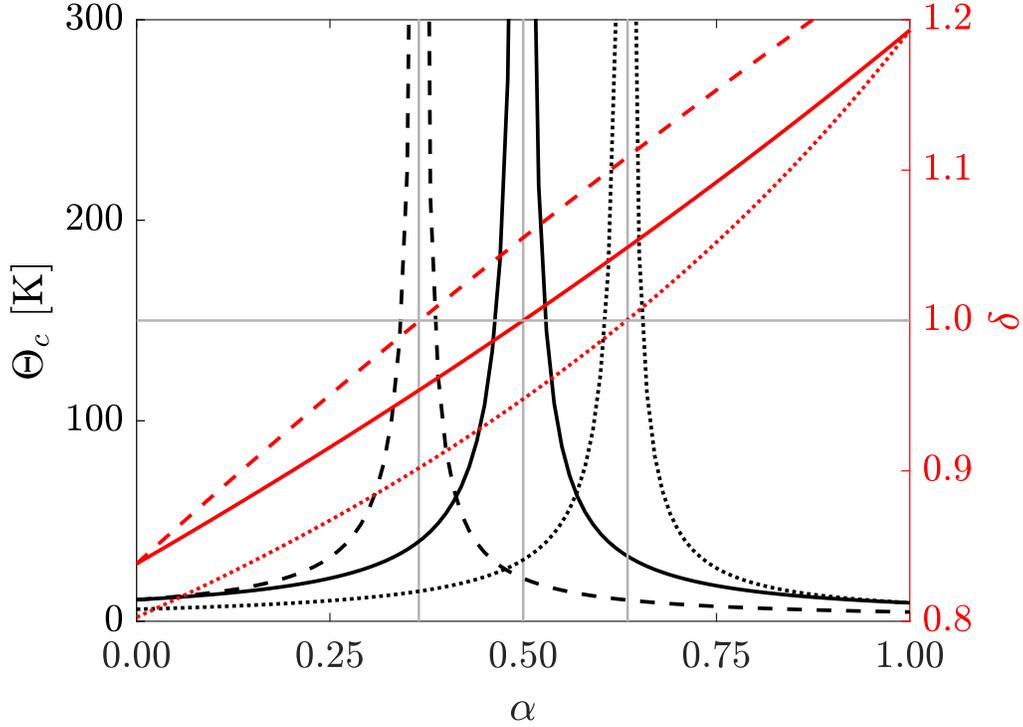}
	\caption{Influence of $\alpha$ on $\Theta_c$ and $\delta$ for cases I (dashed line), II (continuous line), and III (dotted line). The buckling mode is axisymmetric ($\bar{m}=0$, coffee cup) when $\delta<1$ and non-axisymmetric of period $\pi$ ($\bar{m}=2$, saddle) when $\delta>1$. Here, $R=75$ mm, $H=1.5$ mm. Values of the thermo-elastic parameters per case are given in Table \ref{tab_inputs_delta}.}
	\label{fig_L3_delta}
\end{figure}

The occurrence of buckling is determined by the pre-buckled stresses, which are described by the zeroth-order Airy stress function:
\begin{align}
    \chi^{(0)}(\rho) &= - \frac{H^3}{\bar{\bar{D}}} \left( \frac{R}{H} \right)^2 \frac{k_{2,0} \left( a_0 + b_0 \right) - k_{1,0} \left( b_0 + c_0 \right)}{a_0 - c_0} \left( \frac{\rho^{\lambda+1}}{\lambda+1} - \frac{\rho^2}{2} \right) \bar{\Theta} \nonumber \\
    &= - \frac{H^3}{\bar{\bar{D}}} \left( \frac{R}{H} \right)^2 k_{2,0} \left( 1 + \frac{b_0}{a_0} \right) \frac{1 - \delta^2}{1 - \lambda^2}\left( \frac{\rho^{\lambda+1}}{\lambda+1} - \frac{\rho^2}{2} \right) \bar{\Theta}~, \label{eq_chi0}
\end{align}
as follows from Equation (\ref{eq_chi0_full}) with $\mu=\mu(\bar{\Theta})$ according to Equation (\ref{eq_def_mu}), while the parameters $\lambda$ and $\delta$ were introduced in Equation (\ref{eq_def_lambda_delta}). When the material properties are such that $\chi^{(0)} \equiv 0$, the disk behaves macroscopically as if it were isotropic and no stress is built-up during cooling. In our previous work, the condition for no buckling was formulated in terms of $\omega$ (ratio of elastic moduli) and $\gamma$ (ratio of the coefficients of thermal expansion) for uniform orthotropic disks. Their values follow the same trends with respect to fiber orientation: $\omega,\gamma<1$ for a radial alignment, $\omega,\gamma>1$ for a tangential alignment, $\omega=\gamma=1$ for an isotropic distribution. Therefore, the two parameters are interchangeable in discussing the occurrence of buckling. When multiple layers are considered, $\lambda$ (generalization of $\omega$) and $\delta$ (generalization of $\gamma$) become independent, meaning that $\lambda=1$ does not imply $\delta=1$ and vice versa. It can be seen from Equation (\ref{eq_chi0}) that the stress field does vanish for $\delta \to 1^\pm$, irrespective of the value of $\lambda$. Therefore, $\delta$ is identified here as the relevant parameter to describe the occurrence of buckling. Since its definition involves all the thermo-elastic constants and their geometrical averages, the overall isotropic behavior is thus the result of all these factors combined. Consequently, if one were to control the degree of orientation in the layers or their thickness, the macroscopic isotropic state would be reached and buckling would be avoided. This is numerically shown in Figure \ref{fig_L3_delta}, where all the thermo-elastic properties are kept constant (values provided in Table \ref{tab_inputs_delta}) and the thickness ratio $\alpha$ is varied from 0 to 1. A predominant tangential fiber orientation is assumed in the core and a radial one in the skin layer. Three cases are analyzed: in case I, the orientation is stronger in the core, and conversely stronger in the skin in case III; case II describes a balanced configuration, with the same degree of orientation in all layers. The buckling temperature $\Theta_c$ is plotted as a function of $\alpha$. The value $\alpha_\text{iso}$ of the thickness ratio for which no buckling occurs (\textit{isotropic thickness ratio}) can be identified as the asymptote in the buckling temperature. On the second axis, the corresponding value of $\delta$ is displayed. As expected, $\delta \to 1$ when the isotropic thickness ratio is approached.\\
The equivalence of ($\omega,\gamma$) and $\delta$ extends to the determination of the buckling mode too. In the uniform case, axisymmetric buckling (\textit{coffee-cup}) was observed for $\omega,\gamma<1$, whereas non-axisymmetric buckling of period $\pi$ (\textit{saddle}) occurred for $\omega,\gamma>1$. Analogous results hold for the skin-core-skin formulation: buckling is axisymmetric for $\alpha<\alpha_\text{iso}$ (when $\delta<1$) and non-axisymmetric of period $\pi$ for $\alpha>\alpha_\text{iso}$ (when $\delta>1$). These results hold irrespective of the value of $\lambda$.\\
The analogy between the uniform and multi-layer geometries and the results discussed above are summarized in Table \ref{tab_1L-3L}. 

\subsection{Comparison with single-layer model}
\begin{figure}
    \centering
    \includegraphics[width=\linewidth]{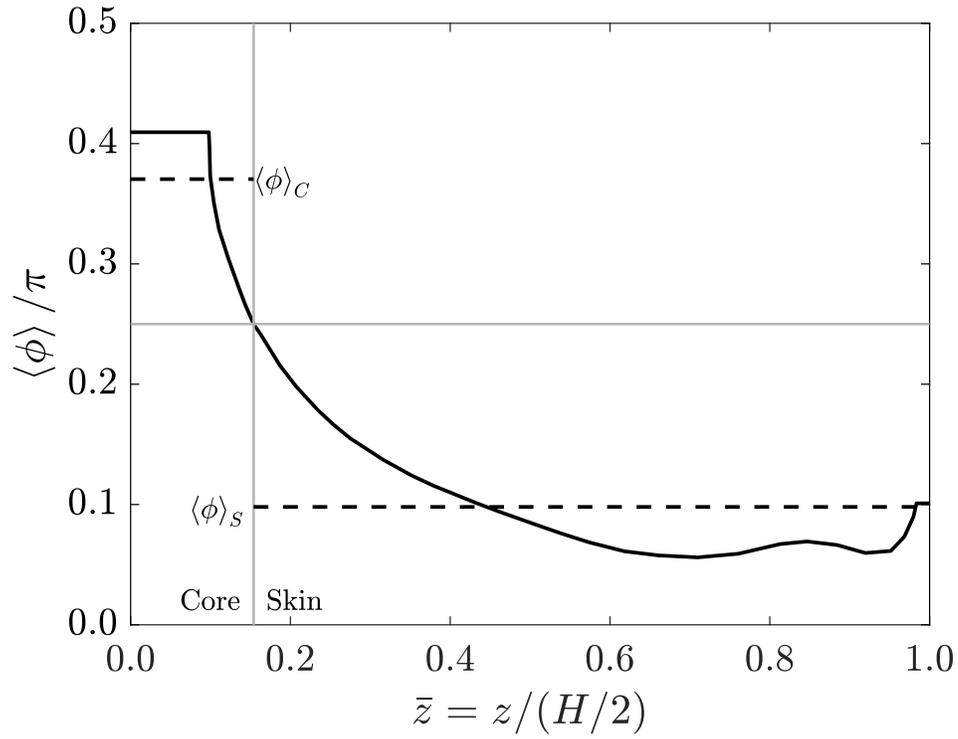}
	\caption{Layer averaged orientation angle for disk of thickness $H=1.5$ mm from \cite{Kikuchi1996a}. Missing data close to the disk center and surface has been replaced by a constant extrapolation. Dashed lines correspond to the averages of $\left<\phi\right>$ over the core and skin layers ($\alpha=0.154$).}
	\label{fig_phi}
\end{figure}

In \cite{Kikuchi1994}, the type and period of buckling was linked to the anisotropy in coefficients of thermal expansion $R\alpha=k_r/k_\varphi$. Experiments were performed by injection-molding disks of PA66 with various reinforcements and the macroscopic orthotropic thermo-elastic properties were measured. In our previous work, it was demonstrated that the proposed model for uniform orthotropic disks is able to correctly capture the buckling mode and excellent quantitative agreement with the results of linear thermo-elastic FEM simulations was shown.\\
Additional simulations were performed in \cite{Kikuchi1996a} to investigate the influence of the disk thickness on the buckling temperature. The fiber orientation distribution was first derived by solving the flow field in the mold and then used as input for the simulations. The values of the thermal parameters averaged over the thickness of the disks were provided, together with the elastic constants corresponding to the highest degree of orientation. The calculations showed that the buckling temperature increases with the disk thickness. The thinnest disks (1.5 to 3 mm) buckled axisymmetrically (coffee cup), whereas the thicker disks (4 and 5 mm) did not show any warpage, as their buckling threshold was higher than the applied temperature difference. These results were already analyzed in our previous work with the model for uniform disks (single-layer formulation) by using the measured thickness-averaged coefficients of linear thermal expansion. For the elastic moduli and Poisson's ratios, the values corresponding to the highest degree of orientation were used. Since the actual material properties vary through the thickness and show less orientation than in the ideal case of perfect alignment, the buckling temperature was underestimated and the deflection was overestimated. The fiber orientation distribution through the disk thickness was studied for the thinnest (1.5 mm) and thickest (5 mm) disks. Since no buckling was predicted for the thickest disk, we focus on the thinnest one. As explained in the next paragraphs, we will translate the fiber orientation distribution to an orientation tensor for a three-layer geometry. The tensor will then be used to estimate the thickness-dependent elastic parameters needed as inputs for the present model.\\
The fiber alignment is described in \cite{Kikuchi1996a} by the \textit{layer averaged orientation angle} $\left<\phi\right>$, defined as the volume average per layer of the fiber angle with respect to the radial direction. When $\left<\phi\right>=0$, all the fibers in the layer are radially oriented, whereas when $\left<\phi\right>=\pi/2$ they are all in the tangential direction. In Figure \ref{fig_phi}, $\left<\phi\right>$ is plotted against the normalized thickness $\bar{z}=z/(H/2)$ for the 1.5mm-thick disk. As $\left<\phi\right>=\pi/4$ corresponds to an average fiber orientation of 45$^\circ$, approximately 15.4\% of the disk shows a predominantly tangential fiber alignment and the remaining 84.6\% a radial one. This implies that in the corresponding skin-core-skin model one should use $\alpha=0.154$. To better describe the fiber orientation distribution in each layer, additional information about the distribution of the angles around their averages should be considered. However, in the following we will assume that all the fibers in a layer share the same orientation angle $\left<\phi\right>$. This is done in order to maintain the focus on the validation of the proposed model. As it is demonstrated in \cite{Advani1987}, the symmetric orientation tensor in cylindrical coordinates for a planar uniform fiber orientation distribution with angle $\theta$ with respect to the radial direction has elements $\alpha_{ij}$ given by
\begin{equation}
    \alpha_{11} = \cos^2\theta; \quad \alpha_{22} = \sin^2\theta; \quad \alpha_{12} = 0. \label{eq_alphaij}
\end{equation}
Let $\left<\phi\right>_S$ and $\left<\phi\right>_C$ be the averages of $\left<\phi\right>$ over the skin and core layers, respectively (see Figure \ref{fig_phi}). By assuming that the orientation of all fibers is given by either of the two averages, the orientation tensor in each layer can be computed directly from Equation (\ref{eq_alphaij}) by taking either $\theta=\left<\phi\right>_S$ or $\theta=\left<\phi\right>_C$.\\
Once the orientation tensor has been estimated, the thermo-elastic constants can be calculated. This is done numerically in Digimat-MF by providing the matrix and fiber thermo-elastic properties, the filler content, and the orientation tensor. Linear thermo-elastic behavior is assumed. Mori-Tanaka homogenization is used in the software to calculate the resulting composite behavior. The inputs used in these calculations are summarized in Table \ref{tab_inputs_digimat} and are extracted from \cite{Kikuchi1996a} and material datasheets, unless otherwise specified. The outcome of this procedure is shown in Table \ref{tab_inputs_valid}. The estimated constants are used as inputs for the three-layer model with $\alpha=0.154$. In agreement with previous results, the model correctly predicts axisymmetric buckling. The buckling threshold and magnitude of warpage at the disk edge are reported in Table \ref{tab_results_valid} and compared to the FEM results and our previous single-layer analysis. Despite the approximation introduced in the estimation of the thermo-elastic parameters, the skin-core-skin geometry significantly improves the quantitative agreement of the predicted buckling temperature difference with the FEM results. The magnitude of the deflection is underpredicted, which is believed to be caused by the conservative calculation of the coefficients of thermal expansion for the composite. A better match can be obtained by improving the calculation of the composite properties, for which more information about the fiber orientation distribution is required.\\

\subsection{Buckling temperature, thickness, and anisotropy}
\begin{figure}
    \centering
    \includegraphics[width=\linewidth]{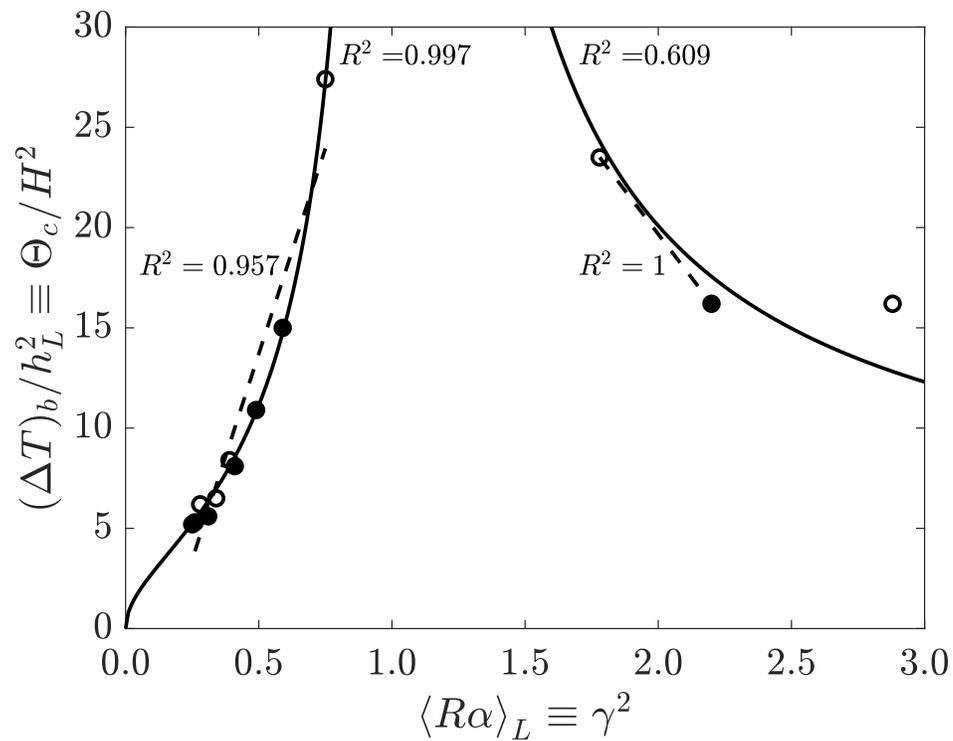}
	\caption{Dependence of the normalized buckling temperature difference on the thermal anisotropy ratio. Experimental data and linear fits are from \cite{Kikuchi1996b}. Solid markers: $h_L=1.5$ mm; empty markers: $h_L=5$ mm. Solid lines correspond to the proposed fit of Equation (\ref{eq_fit}). For $\left< R \alpha \right>_L > 1$, the linear fit is performed on the first two data points only. $R^2$ denotes the statistical coefficient of determination.}
	\label{fig_fit_paper}
\end{figure}

The dependence of the buckling temperature difference $(\Delta T)_b$ on the disk thickness $h$ ($\Theta_c$ and $H$ in our notation, respectively) and on the average thermal anisotropy ratio $\left< R\alpha \right>$ is discussed and quantified in \cite{Kikuchi1996b} by means of FEM simulations for disks of varying thickness and fiber orientation distribution.\\
To investigate the thickness dependence, the fiber orientation distribution was kept fixed while $h$ was varied. It was observed that the ratio $(\Delta T)_b/h^2$ is constant. This property is typical of composite laminates, where the laminar thickness can be changed freely without affecting the thermo-elastic properties. Conversely, injection-molded disks of different thickness will also show a different fiber orientation distribution and hence material properties. In the following, we will thus focus on composite laminates.\\
In our formulation, the buckling temperature for a laminate of total thickness $h_L$ is given by Equation (\ref{eq_def_mu}) (after some rearrangements):
\begin{equation}
    \frac{(\Delta T)_b}{h_L^2} \equiv \frac{\Theta_c}{H^2} = \frac{\bar{\bar{D}}}{H^3} \frac{1}{R^2} \frac{a_0 - c_0}{k_{2,0} \left( a_0 + b_0 \right) - k_{1,0} \left( b_0 + c_0 \right)} \mu_c~.
    \label{eq_def_thetac}
\end{equation}
All the terms on the right including $\mu_c$ depend on the material parameters and relative layer thickness only: the integrals involved in the definitions are already normalized by the appropriate power of $H$ and the ratio $\bar{\bar{D}}/H^3$ is independent of $H$ too. This confirms the numerical results of \cite{Kikuchi1996b}. The previous equation also highlights the dependence on the disk radius $R$, as $\Theta_c \cdot R^2$ is independent of $R$.\\
The second part of \cite{Kikuchi1996b} focuses on the relation between $(\Delta T)_b/h_L^2$ and the average thermal anisotropy ratio $\left< R\alpha \right>_L$. By performing FEM simulations of disks with various thicknesses and fiber orientation distributions, it was shown that linear correlations exist between $(\Delta T)_b/h_L^2$ and $\left< R \alpha \right>_L$ in the intervals $0.26 \leq \left< R \alpha \right>_L \leq 0.75$ and $1.78 \leq \left< R \alpha \right>_L \leq 2.20$, separately. This is shown in Figure \ref{fig_fit_paper} together with the R-squared values for each fit. The two linear functions where combined in the following general expression:
\begin{equation}
    \frac{(\Delta T)_b}{h_L^2} = a \left< R \alpha \right>_L + b + c~\exp \left( \frac{1}{\left|\left< R \alpha \right>_L-1\right|} \right)~.
    \label{eq_fit_lin}
\end{equation}
The values of the coefficients $a$, $b$, and $c$ in each $\left< R \alpha \right>_L$ interval are given in the original publication. The exponential term was introduced to reproduce the divergent behavior of the buckling temperature around $\left< R \alpha \right>_L = 1$. We will now show that our formulation allows to derive analytically a different relation which improves the fit to the experimental data and includes the asymptotic behavior near $\left< R \alpha \right>_L = 1$.\\
Since the attention is put on the average thermal anisotropy ratio, the disk can be analyzed with our previous single-layer formulation, where the ratio of the coefficients of thermal expansion was indicated by $\gamma^2 \equiv \left< R \alpha \right>_L$. Evaluation of Equation (\ref{eq_def_thetac}) for a single layer yields:
\begin{equation}
    \frac{(\Delta T)_b}{h_L^2} = C\left( \omega, \tilde{\nu}, \tilde{k}, R \right) \frac{\sqrt{\left< R \alpha \right>_L}}{1 - \left< R \alpha \right>_L}~,
\end{equation}
where $C$ is a function of the ratio of the elastic moduli $\omega$, the geometric average of the Poisson's ratios $\tilde{\nu}$ and of the coefficients of thermal expansion $\tilde{k}$, and of the disk radius $R$. In principle, the value of $C$ depends on the considered data point of Figure \ref{fig_fit_paper}, as each simulation refers to a different fiber orientation distribution. This is evident when comparing data points for $\left< R \alpha \right>_L < 1$ and $\left< R \alpha \right>_L > 1$: $(\Delta T)_b/h_L^2$ should be always positive, implying that in the first case $C>0$ whereas in the second $C<0$. The lack of detailed information about the thermo-elastic parameters in each simulation does not allow to better quantify the $C$ dependence on them (which is also partially hidden in the numerical calculation of $\mu_c$). However, let us assume that the dependence is weak in the two regions $\left< R \alpha \right>_L < 1$ and $\left< R \alpha \right>_L > 1$ separately, so that the following approximation holds:
\begin{equation}
    \frac{(\Delta T)_b}{h_L^2} \sim 
    \begin{dcases}
        C_1 \sqrt{\left< R \alpha \right>_L} / \left( 1 - \left< R \alpha \right>_L \right)~, \quad \left< R \alpha \right>_L<1~;\\
        C_2 \sqrt{\left< R \alpha \right>_L} / \left( \left< R \alpha \right>_L - 1 \right)~, \quad \left< R \alpha \right>_L>1~.
    \end{dcases}
    \label{eq_fit}
\end{equation}
The constants $C_1$ and $C_2$ are positive and they are determined by fitting the experimental data of Figure \ref{fig_fit_paper}, yielding $C_1=7.88$ and $C_2=14.21$. The fitted function is plotted on the same figure together with its statistical coefficient of determination R-squared. When $\left< R \alpha \right>_L < 1$, the fit to the FEM results is excellent, with an R-squared value higher than that of the linear fit. It can thus be concluded that the dependence of $C$ on the thermo-elastic constants is weak for $\left< R \alpha \right>_L < 1$ and the proposed model of Equation (\ref{eq_fit}) is accurate. When $\left< R \alpha \right>_L > 1$, the R-squared value drops from $0.997$ to $0.609$, indicating that the dependence of $C$ on the thermo-elastic parameters might be stronger than for $\left< R \alpha \right>_L < 1$. As mentioned earlier, the statement cannot be further validated analytically as $\mu_c$ is the solution of a generalized eigenvalue problem and scarce information about the fiber orientation distribution for each simulation is provided. Nevertheless, the proposed fit extends the linear one of Equation (\ref{eq_fit_lin}) to the whole interval $\left< R \alpha \right>_L > 1$, even though more data is needed to better determine its accuracy. The divergent behavior for $\left< R \alpha \right>_L = 1$ is already incorporated in the choice of the fitting function in Equation (\ref{eq_fit}) and does not need to be accounted for with the introduction of corrective terms.
\section{Conclusions}
The thermal buckling of thin injection-molded FRP plates has been investigated via an analytical approach based on the F\"oppl-von K\'arm\'an theory. The presence of short glass fibers is modeled by using cylindrical orthotropy with thermo-elastic material parameters varying through the disk thickness. The formulation was further specified for a skin-core-skin geometry. It was shown that for injection-molded disks there exists a value $\alpha_\text{iso}$ of the core-to-total thickness ratio $\alpha$ for which the disk can be considered macroscopically isotropic, i.e., no buckling occurs. The period of the first buckling mode is determined by the predominant fiber orientation as described by the generalized thermal anisotropy ratio $\delta$: when $\delta<1$ (predominant radial fiber orientation), buckling is axisymmetric (coffee cup); when $\delta>1$ (predominant tangential fiber orientation), buckling is non-axisymmetric of period $\pi$ (saddle); when $\delta=1$, also $\alpha=\alpha_\text{iso}$ and no buckling occurs. This result generalizes that obtained for uniform disks, where the same description holds by replacing $\delta$ with the elastic and thermal anisotropy ratios $\omega$ and $\gamma$, respectively. The skin-core-skin formulation was used for model validation on a case previously analyzed with a uniform fiber orientation distribution. Improvements in the accuracy of the predictions were obtained, showing the importance of including a more detailed description of the fiber orientation distribution through the disk thickness. Based on analytical considerations, the relations between buckling temperature, disk thickness, and thermal anisotropy were explored. An expression correlating the normalized buckling temperature $(\Delta T)_b / h_L^2$ to the thermal anisotropy ratio $\left< R \alpha \right>_L$ was proposed, improving the modified linear fit previously proposed in the literature.
\section{Acknowledgements}
This work was conducted at the DSM Material Science Center research facility in Geleen, the Netherlands. The support of DSM, and especially of Dr. Lucien Douven for the thermo-elastic calculations with Digimat-MF, is highly appreciated. This research did not receive any specific grant from funding agencies in the public, commercial, or not-for-profit sectors.

 \clearpage
\begin{table*}
    \centering
    \begin{tabular}{c|c|ccccccc}
        Case & Layer & \begin{tabular}{@{}c@{}}$E_r$ \\ $[$GPa$]$\end{tabular} & \begin{tabular}{@{}c@{}}$E_\varphi$ \\ $[$GPa$]$\end{tabular} & $\nu_{\varphi r}$ & $\nu_{r \varphi}$ & \begin{tabular}{@{}c@{}}$k_r$ \\ $[$K$^{-1}]$\end{tabular} & \begin{tabular}{@{}c@{}}$k_\varphi$ \\ $[$K$^{-1}]$\end{tabular} & \begin{tabular}{@{}c@{}}$G$ \\ $[$GPa$]$\end{tabular} \\ 
        \hline
        I & \begin{tabular}{@{}c@{}}Core \\ Skin\end{tabular} & \begin{tabular}{@{}c@{}}0.8 \\ 1.0\end{tabular} & \begin{tabular}{@{}c@{}}1.2 \\ 0.8\end{tabular} & \begin{tabular}{@{}c@{}}0.45 \\ 0.24\end{tabular} & \begin{tabular}{@{}c@{}}0.30 \\ 0.30\end{tabular} & \begin{tabular}{@{}c@{}}5.6$\times 10^{-5}$ \\ 2.1$\times 10^{-5}$\end{tabular} & \begin{tabular}{@{}c@{}}2.1$\times 10^{-5}$ \\ 3.9$\times 10^{-5}$\end{tabular} & \begin{tabular}{@{}c@{}}0.3 \\ 0.3\end{tabular} \\
        II & \begin{tabular}{@{}c@{}}Core \\ Skin\end{tabular} & \begin{tabular}{@{}c@{}}0.8 \\ 1.0\end{tabular} & \begin{tabular}{@{}c@{}}1.0 \\ 0.8\end{tabular} & \begin{tabular}{@{}c@{}}0.30 \\ 0.24\end{tabular} & \begin{tabular}{@{}c@{}}0.24 \\ 0.30\end{tabular} & \begin{tabular}{@{}c@{}}3.9$\times 10^{-5}$ \\ 2.1$\times 10^{-5}$\end{tabular} & \begin{tabular}{@{}c@{}}2.1$\times 10^{-5}$ \\ 3.9$\times 10^{-5}$\end{tabular} & \begin{tabular}{@{}c@{}}0.3 \\ 0.3\end{tabular} \\
        III & \begin{tabular}{@{}c@{}}Core \\ Skin\end{tabular} & \begin{tabular}{@{}c@{}}0.8 \\ 1.2\end{tabular} & \begin{tabular}{@{}c@{}}1.0 \\ 0.8\end{tabular} & \begin{tabular}{@{}c@{}}0.30 \\ 0.30\end{tabular} & \begin{tabular}{@{}c@{}}0.24 \\ 0.45\end{tabular} & \begin{tabular}{@{}c@{}}3.9$\times 10^{-5}$ \\ 2.1$\times 10^{-5}$\end{tabular} & \begin{tabular}{@{}c@{}}2.1$\times 10^{-5}$ \\ 5.6$\times 10^{-5}$\end{tabular} & \begin{tabular}{@{}c@{}}0.3 \\ 0.3\end{tabular} 
    \end{tabular}
    \caption{Thermoelastic properties for the cases of Figure \ref{fig_L3_delta}.}
    \label{tab_inputs_delta}
\end{table*}

\begin{table*}
    \centering
	\begin{tabular}{c|cc}
	    Buckling mode & Single-layer & Multi-layer \\
	    \hline
        No buckling & $\omega,\gamma=1$ & $\delta=1$ \\
        Cup ($\bar{m}=0$) & $\omega,\gamma<1$ & $\delta<1$ \\
        Saddle ($\bar{m}=2$) & $\omega,\gamma>1$ & $\delta>1$
    \end{tabular}
	\caption{Equivalence of parameters in the determination of the first buckling mode. Single-layer results from \cite{Gualdi2021}; multi-layer results from present work.}
    \label{tab_1L-3L}
\end{table*}

\begin{table*}
    \centering
	\begin{tabular}{l|cc}
	    & Matrix & Glass fiber \\
	    \hline
	    Elastic modulus & 3.3 GPa & 72.0 GPa \\
	    Poisson's ratio & 0.30 & 0.22 \\
	    CLTE & 8.1$\times 10^{-5}$ K$^{-1}$ & 0.5$\times 10^{-5}$ K$^{-1}$ 
    \end{tabular}
	\caption{Inputs used for Digimat calculations. Filler content: 33 wt\%; Fiber aspect ratio: 20. Matrix elastic modulus and fiber aspect ratio are determined by calculating thermo-elastic composite properties for unidirectional fiber alignment and matching them with the values for the highest degree of orientation reported in \cite{Kikuchi1996a}.}
    \label{tab_inputs_digimat}
\end{table*}

\begin{table*}
    \centering
	\begin{tabular}{c|cccccccccc}
	    & $\left<\phi\right>/\pi$ & $\alpha_{11}$ & $\alpha_{22}$ & \begin{tabular}{@{}c@{}}$E_r$ \\ $[$GPa$]$\end{tabular} & \begin{tabular}{@{}c@{}}$E_\varphi$ \\ $[$GPa$]$\end{tabular} & $\nu_{\varphi r}$ & $\nu_{r \varphi}$ & \begin{tabular}{@{}c@{}}$k_r$ \\ $[$K$^{-1}]$\end{tabular} & \begin{tabular}{@{}c@{}}$k_\varphi$ \\ $[$K$^{-1}]$\end{tabular} & \begin{tabular}{@{}c@{}}$G$ \\ $[$GPa$]$\end{tabular} \\
	    \hline
        Core & 0.37 & 0.16 & 0.84 & 5.26 & 10.91 & 0.35 & 0.17 & 6.18 $\times 10^{-5}$ & 2.37 $\times 10^{-5}$ & 2.35 \\
        Skin & 0.10 & 0.91 & 0.09 & 11.86 & 4.95 & 0.14 & 0.27 & 2.24 $\times 10^{-5}$ & 6.79 $\times 10^{-5}$ & 2.13 \\
        \hline
        Highest & 0.00 & 1.00 & 0.00 & 13.17 & 4.60 & 0.10 & 0.29 & 2.15 $\times 10^{-5}$ & 7.61 $\times 10^{-5}$ & 1.78
    \end{tabular}
	\caption{Estimated thermo-elastic constants for skin-core-skin model based on layer information provided in \cite{Kikuchi1996a}. The last row corresponds to a uniform fiber alignment in the radial direction (\textit{highest degree of orientation} in \cite{Kikuchi1996a}). The calculation is in good agreement with the reported values with the exception of the coefficients of thermal expansion, which are slightly underestimated.}
    \label{tab_inputs_valid}
\end{table*}

\begin{table*}
    \centering
	\begin{tabular}{c|ccccc}
	    Model & $\bar{m}$ & \begin{tabular}{@{}c@{}}$\Theta_c$ \\ $[$K$]$\end{tabular} & err$_\Theta$ & \begin{tabular}{@{}c@{}}$w(1,0)$ \\ $[$mm$]$\end{tabular} & err$_w$ \\
	    \hline
        \cite{Kikuchi1996a} & 0 & 21.00 & - & 7.86 & - \\
        \cite{Gualdi2021} & 0 & 16.89 & -19.57\% & 8.59 & 9.29\% \\
        Present & 0 & 20.91 & -0.43\% & 7.03 & -10.56\%
    \end{tabular}
	\caption{Comparison between the results of \cite{Kikuchi1996a} and our model in its single- and multi-layer formulation. $\bar{m}$ identifies the buckling mode; $\Theta_c$ is the buckling temperature difference; $w(1,0)$ is the deflection at the edge of the disk. Errors are defined as $\text{err}=(\text{model value}-\text{FEM value})/\text{FEM value} \times 100\%$.}
    \label{tab_results_valid}
\end{table*}

 \clearpage

\appendix

\setcounter{section}{0}
\renewcommand{\thesection}{\Alph{section}}
\section{Appendix A}
\setcounter{equation}{0}
\renewcommand{\theequation}{\thesection.\arabic{equation}}
\label{app_A}

\emph{All the results in this Appendix are obtained analogous to \cite{Gualdi2021}.}

\subsection{Differential operators in polar coordinates}
In the dimensionless F\"oppl-von K\'arm\'an equations (\ref{eq_fvk_system}) for orthotropic disks we have
\begin{align}
    \tilde{\Delta}_1^2 f &= \frac{H^3}{\bar{\bar{D}}} \left[ a_2 \left( \frac{\partial^4 f}{\partial \rho^4} + \frac{2}{\rho}\frac{\partial^3 f}{\partial \rho^3} \right) \right. \nonumber \\
    &+ 2 \left( b_2 + 2 d_2 \right) \left( \frac{1}{\rho^4}\frac{\partial^2 f}{\partial \varphi^2} - \frac{1}{\rho^3}\frac{\partial^3 f}{\partial \rho \partial \varphi^2} + \frac{1}{\rho^2}\frac{\partial^4 f}{\partial \rho^2 \partial \varphi^2} \right)  \nonumber\\
    &\left. + c_2 \left( \frac{1}{\rho^4}\frac{\partial^4 f}{\partial \varphi^4} - \frac{1}{\rho^2}\frac{\partial^2 f}{\partial \rho^2} + \frac{2}{\rho^4}\frac{\partial^2 f}{\partial \varphi^2} + \frac{1}{\rho^3}\frac{\partial f}{\partial \rho} \right) \right]~;
\end{align}
\begin{align}
    \tilde{\Delta}_2^2 f &= \bar{\bar{E}} \left[ \frac{1}{c_0 \eta} \left( \frac{\partial^4 f}{\partial \rho^4} + \frac{2}{\rho}\frac{\partial^3 f}{\partial \rho^3} \right) \right. \nonumber \\
    &+ \left( \frac{1}{d_0} -  \frac{2b_0}{a_0 c_0 \eta} \right) \left( \frac{1}{\rho^4}\frac{\partial^2 f}{\partial \varphi^2} - \frac{1}{\rho^3}\frac{\partial^3 f}{\partial \rho \partial \varphi^2} + \frac{1}{\rho^2}\frac{\partial^4 f}{\partial \rho^2 \partial \varphi^2} \right) \nonumber \\
    &\left. + \frac{1}{a_0 \eta} \left( \frac{1}{\rho^4}\frac{\partial^4 f}{\partial \varphi^4} - \frac{1}{\rho^2}\frac{\partial^2 f}{\partial \rho^2} + \frac{2}{\rho^4}\frac{\partial^2 f}{\partial \varphi^2} + \frac{1}{\rho^3}\frac{\partial f}{\partial \rho} \right) \right]~,
\end{align}
with $\eta = 1-b_0^2/(a_0 c_0)$;
\begin{align}
    \left[ f, g \right] &= -\frac{2}{\rho^4}\frac{\partial f}{\partial \varphi}\frac{\partial g}{\partial \varphi} + \frac{2}{\rho^3}\frac{\partial}{\partial \rho}\left( \frac{\partial f}{\partial \varphi}\frac{\partial g}{\partial \varphi}\right) \nonumber \\
    &+ \frac{1}{\rho^2}\left( \frac{\partial^2 f}{\partial \rho^2}\frac{\partial^2 g}{\partial \varphi^2} - 2\frac{\partial^2 f}{\partial \rho \partial \varphi}\frac{\partial^2 g}{\partial \rho \partial \varphi} + \frac{\partial^2 f}{\partial \varphi^2}\frac{\partial^2 g}{\partial \rho^2}\right) + \frac{1}{\rho}\frac{\partial}{\partial \rho}\left( \frac{\partial f}{\partial \rho}\frac{\partial g}{\partial \rho}\right)~,
\end{align}
where $\bar{\bar{E}}$ and $\bar{\bar{D}}$ are defined in Equation (\ref{eq_def_barbars}), and
\begin{align}
    &a_0 = \frac{1}{H} \int_{-H/2}^{H/2} \frac{\tilde{E}(z)}{1-\tilde{\nu}(z)^2}~\frac{1}{\omega(z)} \, \mathrm{d}z ~, \quad &&a_2 = \frac{1}{H^3} \int_{-H/2}^{H/2} \frac{\tilde{E}(z) z^2}{1-\tilde{\nu}(z)^2}~\frac{1}{\omega(z)} \, \mathrm{d}z ~;\\
    &b_0 = \frac{1}{H} \int_{-H/2}^{H/2} \frac{\tilde{E}(z)}{1-\tilde{\nu}(z)^2}~\tilde{\nu}(z) \, \mathrm{d}z ~, &&b_2 = \frac{1}{H^3} \int_{-H/2}^{H/2} \frac{\tilde{E}(z) z^2}{1-\tilde{\nu}(z)^2}~\tilde{\nu}(z) \, \mathrm{d}z ~;\\
    &c_0 = \frac{1}{H} \int_{-H/2}^{H/2} \frac{\tilde{E}(z)}{1-\tilde{\nu}(z)^2}~\omega(z) \, \mathrm{d}z ~, &&c_2 = \frac{1}{H^3} \int_{-H/2}^{H/2} \frac{\tilde{E}(z) z^2}{1-\tilde{\nu}(z)^2}~\omega(z) \, \mathrm{d}z ~;\\
    &d_0 = \frac{1}{H} \int_{-H/2}^{H/2} G(z) \, \mathrm{d}z ~, &&d_2 = \frac{1}{H^3} \int_{-H/2}^{H/2} G(z) z^2 \, \mathrm{d}z ~.
\end{align}
Moreover, for the dimensionless buckling parameter $\mu$ as defined in Equation (\ref{eq_def_mu}), we need
\begin{align}
    k_{1,0} &= \frac{1}{H}~\int_{-H/2}^{H/2} \frac{\tilde{E}(z) \tilde{k}(z)}{1-\tilde{\nu}(z)^2} \left[ \frac{\gamma(z)}{\omega(z)} + \frac{\tilde{\nu}(z)}{\gamma(z)} \right] \, \mathrm{d}z~; \\
    k_{2,0} &= \frac{1}{H}~\int_{-H/2}^{H/2} \frac{\tilde{E}(z) \tilde{k}(z)}{1-\tilde{\nu}(z)^2} \left[ \tilde{\nu}(z)\gamma(z) + \frac{\omega(z)}{\gamma(z)} \right] \, \mathrm{d}z~.
 \end{align}

\subsection{Boundary conditions}
For a free disk at $\rho=1$, $N_n=N_s=M_b=V_v=0$, yielding successively:
\begin{align}
    &\frac{1}{\rho} \frac{\partial \chi}{\partial \rho} + \frac{1}{\rho^2} \frac{\partial^2 \chi}{\partial \varphi^2} = 0~; \label{eq_bcNn_expl} \\
    &- \frac{1}{\rho^2} \frac{\partial \chi}{\partial \varphi} + \frac{1}{\rho} \frac{\partial^2 \chi}{\partial \rho \partial \varphi} = 0~; \label{eq_bcNs_expl} \\
    &a_2 \frac{\partial^2 w}{\partial \rho^2} + b_2 \left( \frac{1}{\rho^2}\frac{\partial^2 w}{\partial \varphi^2} + \frac{1}{\rho}\frac{\partial w}{\partial \rho}\right) \eqqcolon \mathscr{M}\left( w \right) = 0~; \label{eq_bcMb_expl} \\
    &\left( 4 d_2 + b_2 \right) \left( \frac{1}{\rho^3}\frac{\partial^2 w}{\partial \varphi^2} - \frac{1}{\rho^2}\frac{\partial^3 w}{\partial \rho \partial\varphi^2}\right) + c_2 \left( \frac{1}{\rho^3}\frac{\partial^2 w}{\partial \varphi^2} + \frac{1}{\rho^2}\frac{\partial w}{\partial \rho}\right) \nonumber \\
    &\quad - a_2 \left( \frac{\partial^3 w}{\partial \rho^3} + \frac{1}{\rho}\frac{\partial^2 w}{\partial \rho^2} \right) \eqqcolon \mathscr{V}\left( w \right) = 0~. \label{eq_bcVv_expl}
\end{align}
\setcounter{section}{1}
\renewcommand{\thesection}{\Alph{section}}
\section{Appendix B}
\setcounter{equation}{0}
\renewcommand{\theequation}{\thesection.\arabic{equation}}
\label{app_B}

\subsection{Zeroth order}
\noindent\textbf{Problem}

\begin{equation}
    \tilde{\Delta}_2^2 \chi^{(0)} = 0~.
\end{equation}

\noindent\textbf{Solution}

\begin{equation}
    \chi^{(0)} \left( \rho; \mu \right) = - \mu \left( \frac{\rho^{\lambda+1}}{\lambda+1} - \frac{\rho^2}{2}\right) =: - \mu \check{\chi}^{(0)} \left( \rho \right)~, \label{eq_chi0_full}
\end{equation}
with $\lambda$ as defined in Equation (\ref{eq_def_lambda_delta}).

\subsection{First order}
\noindent\textbf{Problem}

\begin{equation}
    \tilde{\Delta}_1^2 w^{(1)} - \left[ w^{(1)}, \chi^{(0)} \right] = 0~,
\end{equation}
where $w^{(1)}$ is sought for by using a Legendre-Fourier expansion in combination with a Galerkin approach:
\begin{equation}
    w^{(1)}(\rho, \varphi) = \sum_{m=0}^M \sum_{j=0}^{N-2} a_j^{(1)}(m) \Phi_j(2\rho - 1) \cos(m \varphi)~.
\end{equation}

\noindent\textbf{Solution}

The vector of unknown coefficients $\boldsymbol{a}^{(1)}(m) = \left\lbrace a^{(1)}_j(m) \right\rbrace_{j=0,...,N-2}$ is solution of the following linear system with symmetric matrix of coefficients:
\begin{equation}
    \boldsymbol{\mathcal{W}}(m) \boldsymbol{a}^{(1)}(m) = \boldsymbol{0}~, \quad m=0,...,M,
\end{equation}
\begin{equation}
    \boldsymbol{\mathcal{W}}(m) = \boldsymbol{\mathcal{W}}_1(m) - \mu \boldsymbol{\mathcal{W}}_2(m)~,
\end{equation}
where
\begin{align}
    \boldsymbol{\mathcal{W}}_1(m) &= \frac{H^3}{\bar{\bar{D}}} \left\lbrace a_2 \boldsymbol{\mathcal{A}} + \left[ c_2 + 2 m^2 \left( b_2 + 2 d_2 \right) \right] \boldsymbol{\mathcal{B}} +  m^2 \left[ c_2 \left( m^2 - 2 \right) - 2 \left( b_2 + 2 d_2 \right) \right] \boldsymbol{\mathcal{C}} \right. \nonumber \\
    &+ \left. b_2 \boldsymbol{\mathcal{P}} - m^2 \left( c_2 + b_2 + 4 d_2 \right) \boldsymbol{\mathcal{Q}} - m^2 b_2 \boldsymbol{\mathcal{R}} \right\rbrace ~,
\end{align}
and
\begin{equation}
    \boldsymbol{\mathcal{W}}_2(m) = \boldsymbol{\mathcal{D}} + \frac{m^2}{2} \boldsymbol{\mathcal{E}}~,
\end{equation}
with
\begin{align}
    &\boldsymbol{\mathcal{A}} = \left\lbrace a_{ij} \right\rbrace_{i,j}~, \quad &&a_{ij} = \int_{-1}^{1} \! (t+1)\Phi_j'' \Phi_i'' \, \mathrm{d}t~; \label{eq_aij}\\
    &\boldsymbol{\mathcal{B}} = \left\lbrace b_{ij} \right\rbrace_{i,j}~, &&b_{ij} = \int_{-1}^{1} \! \frac{1}{t+1} \Phi_j' \Phi_i' \, \mathrm{d}t~; \label{eq_bij}\\
    &\boldsymbol{\mathcal{C}} = \left\lbrace c_{ij} \right\rbrace_{i,j}~, &&c_{ij} = \int_{-1}^{1} \! \frac{1}{(t+1)^3} \Phi_j \Phi_i \, \mathrm{d}t~; \label{eq_cij}\\
    &\boldsymbol{\mathcal{D}} = \left\lbrace d_{ij} \right\rbrace_{i,j}~, &&d_{ij} = \frac{1}{2}\int_{-1}^{1} \! \Phi_j' \Phi_i' \left[ \left( \frac{t+1}{2}\right) ^{\lambda} - \frac{t+1}{2}\right] \, \mathrm{d}t~; \label{eq_dij}\\
    &\boldsymbol{\mathcal{E}} = \left\lbrace e_{ij} \right\rbrace_{i,j}~, &&e_{ij} = \frac{1}{2}\int_{-1}^{1} \! \frac{1}{t+1} \Phi_j \Phi_i \left[ \lambda\left( \frac{t+1}{2}\right) ^{\lambda-1} - 1 \right] \, \mathrm{d}t~; \label{eq_eij}\\
    &\boldsymbol{\mathcal{P}} = \left\lbrace p_{ij} \right\rbrace_{i,j}~, &&p_{ij} = \Phi_j' \Phi_i' \big|_{t=1}~; \label{eq_pij}\\
    &\boldsymbol{\mathcal{Q}} = \left\lbrace q_{ij} \right\rbrace_{i,j}~, &&q_{ij} = \frac{1}{(t+1)^2} \Phi_j \Phi_i \big|_{t=1}~; \label{eq_qij}\\
    &\boldsymbol{\mathcal{R}} = \left\lbrace r_{ij} \right\rbrace_{i,j}~, &&r_{ij} = \frac{1}{t+1} \left( \Phi_j' \Phi_i + \Phi_j \Phi_i' \right) \big|_{t=1}~. \label{eq_rij}
\end{align}
The approximate first-order deflection $w_N^{(1)}$ is determined up to an unknown multiplicative constant $A^{(1)}$:
\begin{equation}
    w_N^{(1)}(\rho, \varphi) = A^{(1)} \sum_{j=0}^{N-2} a_j^{(1)} \Phi_j(2\rho - 1) \cos(\bar{m} \varphi)~.
\end{equation}

\subsection{Second order}
\noindent\textbf{Problem}

\begin{equation}
    \tilde{\Delta}_2^2 \chi^{(2)} = - \frac{1}{2} \left[ w^{(1)}, w^{(1)} \right]~.
\end{equation}

\noindent\textbf{Solution}

We rewrite $w^{(1)}_N(\rho,\varphi)$ in powers of $\rho$ as:
\begin{equation}
	w^{(1)}_N(\rho,\varphi) = A^{(1)} \sum_{k=2}^N \tilde{a}_k^{(1)} \rho^k \cos \left( \bar{m} \varphi \right)~.
\end{equation}
The second-order Airy potential $\chi^{(2)}_N$ is then expressed by:
\begin{equation}
        \chi^{(2)}_N(\rho,\varphi;\bar{m}) = \chi^{(0)}(\rho)|_{\mu=\mu_c} + \left[ A^{(1)}\right]^2 \left[ \chi^*_0(\rho;\bar{m}) + \chi^*_2(\rho;\bar{m}) \cos(2\bar{m}\varphi) \right]~,
\end{equation}
where
\begin{align}
    \chi^*_0(\rho;m) &= \sum_{j=2}^N \sum_{k=2}^N \frac{j+k}{2} \left( \frac{\rho^{\lambda+1}}{\lambda+1} - \frac{\rho^{j+k}}{j+k}\right) \frac{\tilde{a}^{(1)}_j \tilde{a}^{(1)}_k}{x_0(j,k;m)}~; \label{eq_chistar0} \\
    \chi^*_2(\rho;m) &= \sum_{j=2}^N \sum_{k=2}^N \frac{1}{2} \left[ \Lambda_1(j,k;m)~\rho^{\lambda_1(m)} - \Lambda_2(j,k;m)~\rho^{\lambda_2(m)} - \rho^{j+k} \right] \frac{\tilde{a}^{(1)}_j \tilde{a}^{(1)}_k}{x_2(j,k;m)}~; \\
    \Lambda_1(j,k;m) &= \frac{j+k-\lambda_2(m)}{\lambda_1(m)-\lambda_2(m)}~,~~~~\Lambda_2(j,k;m) = \frac{j+k-\lambda_1(m)}{\lambda_1(m)-\lambda_2(m)}~, \\
    \lambda_1(m) &= 1 + \sqrt{ \beta_1(m) + \sqrt{ \beta_1(m)^2 - \beta_2(m)}}~; \label{eq_chistar2} \\
    \lambda_2(m) &= 1 + \sqrt{ \beta_1(m) - \sqrt{ \beta_1(m)^2 - \beta_2(m)}}~; \\
    \beta_1(m) &= \frac{1}{2} \left[ 1 + \lambda^2 + 4 m^2 c_0 \left( \frac{\eta}{d_0} - \frac{2b_0}{a_0 c_0} \right) \right]~; \\
    \beta_2(m) &= \lambda^2 \left( 4 m^2 - 1 \right)^2~; \\
    x_0(j,k;m) &= \frac{1}{ \bar{\bar{E}}}\frac{(j+k)(j+k-2)(j+k+\lambda-1)(j+k-\lambda-1)}{c_0 \eta \left[ \alpha(j,k;m) - \beta(j,k;m) \right] }~; \\
    x_2(j,k;m) &= \frac{\bar{\bar{E}}} {\eta \left[ \alpha(j,k;m) + \beta(j,k;m) \right]} \left[ \frac{(j+k-1)^2(j+k-2)(j+k)}{c_0} \right. \nonumber \\
    &- \left.  4m^2(j+k-1)^2 \left( \frac{\eta}{d_0} - \frac{2b_0}{a_0 c_0} \right) - \frac{ (j+k-2)(j+k) - 8m^2( 2m^2-1) }{a_0} \right]~;
\end{align}

\subsection{Third order}
\noindent\textbf{Problem}

\begin{equation}
    \tilde{\Delta}_1^2 w^{(3)} - \left[ w^{(3)}, \chi^{(0)} \right] = \left[ w^{(1)}, \chi^{(2)} \right]~,
\end{equation}
where $w^{(3)}$ is sought for by using a Galerkin-Fourier expansion:
\begin{equation}
    w^{(3)}(\rho, \varphi) = \sum_{m=0}^M \sum_{j=0}^{N-2} a_j^{(3)}(m) \Phi_j(2\rho - 1) \cos(m \varphi)~.
\end{equation}

\noindent\textbf{Solution}

The vector of unknown coefficients $\boldsymbol{a}^{(3)}(m) = \left\lbrace a^{(3)}_j(m) \right\rbrace_{j=0,...,N-2}$ is solution of the following linear system with symmetric matrix of coefficients:
\begin{equation}
    \boldsymbol{\mathcal{W}}(m) \boldsymbol{a}^{(3)}(m) = \boldsymbol{\mathcal{F}}(m)
    \begin{bmatrix}
        A^{(1)} \\
        -\left[ A^{(1)}\right] ^3
    \end{bmatrix}~, \\
    \quad m=0,...,M~,
\end{equation}
where $\boldsymbol{\mathcal{W}}(m)$ was already introduced at first order (and is evaluated for $\mu=\mu_c$) and the matrix $\boldsymbol{\mathcal{F}}(m)$ is defined as
\begin{equation}
    \boldsymbol{\mathcal{F}}(m) =
    \begin{bmatrix}
        x(m)_0 & y(m)_0 \\
        \vdots & \vdots \\
        x(m)_{N-2} & y(m)_{N-2}
    \end{bmatrix}~,
\end{equation}
with $(2\rho-1 \rightarrow t)$
\begin{align}
    x(m=\bar{m})_i &= \mu_c \sum_{j=0}^{N-2} a^{(1)}_j \left\lbrace \int_{-1}^{1} \! \frac{\bar{m}^2}{t+1} \left[ \lambda \left( \frac{t+1}{2}\right) ^{\lambda - 1} - 1 \right] \Phi_i \Phi_j \, \mathrm{d}t \right. \nonumber \\
    &+ \left. \int_{-1}^{1} \! 2 \left[ \left( \frac{t+1}{2}\right)^{\lambda} - \frac{t+1}{2} \right] \Phi_i' \Phi_j' \, \mathrm{d}t \right\rbrace~; \label{eq_xm} \\
    y(m=\bar{m})_i &= \sum_{j=0}^{N-2} a^{(1)}_j \left\lbrace \int_{-1}^{1} \! \bar{m}^2 \left[ Y + \frac{4}{t+1}(\chi^*_0)'' \right] \Phi_i \Phi_j \, \mathrm{d}t \right. \nonumber \\
    &+ \left. \int_{-1}^{1} \! 4\left[ \frac{1}{2}(\chi^*_2)' - \frac{2\bar{m}^2}{t+1}\chi^*_2 + (\chi^*_0)' \right] \Phi'_i \Phi'_j \, \mathrm{d}t \right\rbrace~; \label{eq_ym} \\ 
    Y &= \frac{8}{(t+1)^3}\chi^*_2 - \frac{8}{(t+1)^2}(\chi^*_2)' + \frac{2}{t+1}(\chi^*_2)''~; \\
    y(m=3\bar{m})_i &= \sum_{j=0}^{N-2} a^{(1)}_j \left\lbrace -\int_{-1}^{1} \! \bar{m}^2 Y \Phi_i \Phi_j \, \mathrm{d}t \right. \nonumber \\
    &+ \left. \int_{-1}^{1} \! 2\left[ \frac{4\bar{m}^2}{t+1}\chi^*_2 + (\chi^*_2)' \right] \Phi'_i \Phi'_j \, \mathrm{d}t + \int_{-1}^{1} \! \frac{16\bar{m}^2}{t+1}\chi^*_2 \Phi_i \Phi''_j \, \mathrm{d}t \right\rbrace~. \label{eq_y3m}
\end{align}
All the other elements of $\boldsymbol{\mathcal{F}}(m)$ are zeroes. The first-order deflection is fully determined by taking the singular value decomposition of $\boldsymbol{\mathcal{W}}(\bar{m})$:
\begin{equation}
	\boldsymbol{\mathcal{W}}(\bar{m}) = \boldsymbol{\mathcal{U}} \boldsymbol{\Sigma} \boldsymbol{\mathcal{V}}^T~,
\end{equation}
yielding
\begin{equation}
	A^{(1)} = \pm \sqrt{\frac{\left( \boldsymbol{\mathcal{U}}^T \boldsymbol{\mathcal{F}}(\bar{m}) \right)_{N-1,1}}{\left( \boldsymbol{\mathcal{U}}^T \boldsymbol{\mathcal{F}}(\bar{m}) \right)_{N-1,2}}}~.
\end{equation}

\bibliographystyle{model2-names}
\bibliography{mybibfile}

\begin{thebibliography}{14}
\expandafter\ifx\csname natexlab\endcsname\relax\def\natexlab#1{#1}\fi
\providecommand{\url}[1]{\texttt{#1}}
\providecommand{\href}[2]{#2}
\providecommand{\path}[1]{#1}
\providecommand{\DOIprefix}{doi:}
\providecommand{\ArXivprefix}{arXiv:}
\providecommand{\URLprefix}{URL: }
\providecommand{\Pubmedprefix}{pmid:}
\providecommand{\doi}[1]{\href{http://dx.doi.org/#1}{\path{#1}}}
\providecommand{\Pubmed}[1]{\href{pmid:#1}{\path{#1}}}
\providecommand{\bibinfo}[2]{#2}
\ifx\xfnm\relax \def\xfnm[#1]{\unskip,\space#1}\fi
\bibitem[{Adhikari et~al.(2016)Adhikari, Bourgade and Asundi}]{Adhikari2016}
\bibinfo{author}{Adhikari, A.}, \bibinfo{author}{Bourgade, T.},
  \bibinfo{author}{Asundi, A.}, \bibinfo{year}{2016}.
\newblock \bibinfo{title}{Residual stress measurement for injection molded
  components}.
\newblock \bibinfo{journal}{Theoretical and Applied Mechanics Letters}
  \bibinfo{volume}{6}, \bibinfo{pages}{152 -- 156}.
\newblock \DOIprefix\doi{https://doi.org/10.1016/j.taml.2016.04.004}.
\bibitem[{Advani and Tucker(1987)}]{Advani1987}
\bibinfo{author}{Advani, S.}, \bibinfo{author}{Tucker, C.},
  \bibinfo{year}{1987}.
\newblock \bibinfo{title}{The use of tensors to describe and predict fiber
  orientation in short fiber composites}.
\newblock \bibinfo{journal}{Journal of Rheology} \bibinfo{volume}{31},
  \bibinfo{pages}{751--784}.
\newblock \DOIprefix\doi{10.1122/1.549945}.
\bibitem[{Baranowski et~al.(2019)Baranowski, Dobrovolskij, Dremel, Hölzing,
  Lohfink, Schladitz and Zabler}]{Baranowski2019}
\bibinfo{author}{Baranowski, T.}, \bibinfo{author}{Dobrovolskij, D.},
  \bibinfo{author}{Dremel, K.}, \bibinfo{author}{Hölzing, A.},
  \bibinfo{author}{Lohfink, G.}, \bibinfo{author}{Schladitz, K.},
  \bibinfo{author}{Zabler, S.}, \bibinfo{year}{2019}.
\newblock \bibinfo{title}{Local fiber orientation from x-ray region-of-interest
  computed tomography of large fiber reinforced composite components}.
\newblock \bibinfo{journal}{Composites Science and Technology}
  \bibinfo{volume}{183}, \bibinfo{pages}{107786}.
\newblock \DOIprefix\doi{https://doi.org/10.1016/j.compscitech.2019.107786}.
\bibitem[{Gualdi et~al.(2021)Gualdi, {van de Ven} and Slot}]{Gualdi2021}
\bibinfo{author}{Gualdi, A.}, \bibinfo{author}{{van de Ven}, A.A.F.},
  \bibinfo{author}{Slot, J.J.M.}, \bibinfo{year}{2021}.
\newblock \bibinfo{title}{Thermal buckling of thin injection-molded {FRP}
  plates}.
\newblock \bibinfo{journal}{International Journal of Solids and Structures}
  \bibinfo{volume}{219-220}, \bibinfo{pages}{120--133}.
\newblock \DOIprefix\doi{https://doi.org/10.1016/j.ijsolstr.2021.02.015}.
\bibitem[{Hamanaka et~al.(2017)Hamanaka, Yamashita, Nonomura, Nguyen~Thi,
  Wakano and Yokoyama}]{Hamanaka2017}
\bibinfo{author}{Hamanaka, S.}, \bibinfo{author}{Yamashita, K.},
  \bibinfo{author}{Nonomura, C.}, \bibinfo{author}{Nguyen~Thi, T.B.},
  \bibinfo{author}{Wakano, T.}, \bibinfo{author}{Yokoyama, A.},
  \bibinfo{year}{2017}.
\newblock \bibinfo{title}{Measurement of fiber orientation distribution in
  injection-molded composites with high filler content}.
\newblock \bibinfo{journal}{PROCEEDINGS OF PPS-32: The 32nd International
  Conference of the Polymer Processing Society - Conference Papers}
  \bibinfo{volume}{1914}, \bibinfo{pages}{140011}.
\newblock \DOIprefix\doi{10.1063/1.5016776}.
\bibitem[{Hetnarski(2014)}]{Hetnarski2014}
\bibinfo{author}{Hetnarski, R.B.}, \bibinfo{year}{2014}.
\newblock \bibinfo{title}{Encyclopedia of Thermal Stresses}.
\newblock \bibinfo{publisher}{Springer}, \bibinfo{address}{Dordrecht}.
\bibitem[{Kikuchi and Koyama(1994)}]{Kikuchi1994}
\bibinfo{author}{Kikuchi, H.}, \bibinfo{author}{Koyama, K.},
  \bibinfo{year}{1994}.
\newblock \bibinfo{title}{Material anisotropy and warpage of nylon 66
  composites}.
\newblock \bibinfo{journal}{Polymer Engineering \& Science}
  \bibinfo{volume}{34}, \bibinfo{pages}{1411--1418}.
\newblock \DOIprefix\doi{https://doi.org/10.1002/pen.760341808}.
\bibitem[{Kikuchi and Koyama(1996a)}]{Kikuchi1996a}
\bibinfo{author}{Kikuchi, H.}, \bibinfo{author}{Koyama, K.},
  \bibinfo{year}{1996}a.
\newblock \bibinfo{title}{The relation between thickness and warpage in a disk
  injection molded from fiber reinforced {PA}66}.
\newblock \bibinfo{journal}{Polymer Engineering \& Science}
  \bibinfo{volume}{36}, \bibinfo{pages}{1317--1325}.
\newblock \DOIprefix\doi{https://doi.org/10.1002/pen.10526}.
\bibitem[{Kikuchi and Koyama(1996b)}]{Kikuchi1996b}
\bibinfo{author}{Kikuchi, H.}, \bibinfo{author}{Koyama, K.},
  \bibinfo{year}{1996}b.
\newblock \bibinfo{title}{Warpage, anisotropy, and part thickness}.
\newblock \bibinfo{journal}{Polymer Engineering \& Science}
  \bibinfo{volume}{36}, \bibinfo{pages}{1326--1335}.
\newblock \DOIprefix\doi{https://doi.org/10.1002/pen.10527}.
\bibitem[{Lionetto et~al.(2021)Lionetto, Montagna, Natali, {De Pascalis},
  Nacucchi, Caretto and Maffezzoli}]{Lionetto2021}
\bibinfo{author}{Lionetto, F.}, \bibinfo{author}{Montagna, F.},
  \bibinfo{author}{Natali, D.}, \bibinfo{author}{{De Pascalis}, F.},
  \bibinfo{author}{Nacucchi, M.}, \bibinfo{author}{Caretto, F.},
  \bibinfo{author}{Maffezzoli, A.}, \bibinfo{year}{2021}.
\newblock \bibinfo{title}{Correlation between elastic properties and morphology
  in short fiber composites by x-ray computed micro-tomography}.
\newblock \bibinfo{journal}{Composites Part A: Applied Science and
  Manufacturing} \bibinfo{volume}{140}, \bibinfo{pages}{106169}.
\newblock \DOIprefix\doi{https://doi.org/10.1016/j.compositesa.2020.106169}.
\bibitem[{Lopatin and Morozov(2008)}]{Lopatin2008}
\bibinfo{author}{Lopatin, A.}, \bibinfo{author}{Morozov, E.},
  \bibinfo{year}{2008}.
\newblock \bibinfo{title}{Symmetrical facing wrinkling of composite sandwich
  panels}.
\newblock \bibinfo{journal}{Journal of Sandwich Structures \& Materials}
  \bibinfo{volume}{10}, \bibinfo{pages}{475--497}.
\newblock \DOIprefix\doi{10.1177/1099636208097196}.
\bibitem[{Oumer and Mamat(2012)}]{Oumer2012}
\bibinfo{author}{Oumer, A.N.}, \bibinfo{author}{Mamat, O.},
  \bibinfo{year}{2012}.
\newblock \bibinfo{title}{A study of fiber orientation in short
  fiber-reinforced composites with simultaneous mold filling and phase change
  effects}.
\newblock \bibinfo{journal}{Composites Part B: Engineering}
  \bibinfo{volume}{43}, \bibinfo{pages}{1087--1094}.
\newblock \DOIprefix\doi{https://doi.org/10.1016/j.compositesb.2012.01.043}.
\bibitem[{Tseng and Osswald(1994)}]{Tseng1994}
\bibinfo{author}{Tseng, S.C.}, \bibinfo{author}{Osswald, T.A.},
  \bibinfo{year}{1994}.
\newblock \bibinfo{title}{Predicting shrinkage and warpage of fiber-reinforced
  composite parts}.
\newblock \bibinfo{journal}{Polymer Composites} \bibinfo{volume}{15},
  \bibinfo{pages}{270--277}.
\newblock \DOIprefix\doi{https://doi.org/10.1002/pc.750150405}.
\bibitem[{Wankhade and Niyogi(2020)}]{Wankhade2020}
\bibinfo{author}{Wankhade, R.}, \bibinfo{author}{Niyogi, S.},
  \bibinfo{year}{2020}.
\newblock \bibinfo{title}{Buckling analysis of symmetric laminated composite
  plates for various thickness ratios and modes}.
\newblock \bibinfo{journal}{Innovative Infrastructure Solutions}
  \bibinfo{volume}{5}, \bibinfo{pages}{65}.
\newblock \DOIprefix\doi{10.1007/s41062-020-00317-8}.

\end{thebibliography}

\end{document}